\documentclass[a4paper]{jpconf}
\usepackage{graphicx}
\begin{document}
\title{Origin of the X-ray emission in the nuclei of {FR\,Is}}

\author{Qingwen Wu$^{1,2}$, Feng Yuan$^{2}$ and Xinwu Cao$^{2}$}

\address{$^{1}$ International Center for Astrophysics, Korean
Astronomy and Space Science Institute, Daejeon 305348, Republic of
Korean;\\ $^{2}$ Shanghai Astronomical Observatory, Chinese Academy
of Sciences, Shanghai, 200030  China}

\ead{qwwu@shao.ac.cn, fyuan@shao.ac.cn, cxw@shao.ac.cn}

\begin{abstract}
We investigate the X-ray origin in {FR\,Is} using the multi-waveband
high resolution data of eight FR I sources, which have very low
Eddington ratios. We fit their multi-waveband spectrum using a
coupled accretion-jet model. We find that X-ray emission in the
source with the highest $L_{\rm X}$ ($\sim 1.8 \times 10^{-4}L_{\rm
Edd}$) is from the advection-dominated accretion flow (ADAF). Four
sources with moderate $L_{\rm X}$ ($\sim$ several $\times
10^{-6}L_{\rm Edd}$) are complicated. The X-ray emission of one
{FR\,I} is from the jet, and the other three is from the sum of the
jet and ADAF. The X-ray emission in the three least luminous sources
($L_{\rm X}\leq1.0\times 10^{-6}L_{\rm Edd}$) is dominated by the
jet. These results roughly support the predictions of Yuan and Cui
(2005) where they predict that when the X-ray luminosity of the
system is below a critical value, the X-radiation will not be
dominated by the emission from the ADAF any longer, but by the jet.
We also find that the accretion rates in four sources must be higher
than the Bondi rates, which implies that other fuel supply (e.g.,
stellar winds) inside the Bondi radius should be important.
\end{abstract}

\section{Introduction}
   Radio galaxies are usually classified as {FR\,I} or {FR\,II} sources
depending on their radio morphology. {FR\,I} radio galaxies (defined
by edge-darkened radio structure) have lower radio power than
{FR\,II} galaxies (defined by edge-brightened radio structure due to
compact jet terminating hot spots) (Fanaroff \& Riley 1974).
Different explanations of division of {FR\,I} and {FR\,II} radio
galaxies invoke either the interaction of the jet with the ambient
medium or the intrinsic nuclei properties of accretion and jet
formation processes (e.g., Bicknell 1995; Reynolds \textit{et al} 1996a;
Hardcastle \textit{et al} 2007). Accretion mode in low power {FR\,Is} may be
different from that in powerful {FR\,IIs}. There is growing evidence
to suggest that {FR\,Is} type radio galaxy nuclei possess
advection-dominated accretion flow (ADAF; or ``radiative inefficient
accretion flows''; see Narayan \& McMlintock 2008 for a review). In
fact, we now have strong observational evidence that ADAFs may be
powering various types of low-luminosity active galactic nuclei
(LLAGNs; e.g., Yuan 2007; Ho 2008 for reviews), not only {FR\,Is}.
It was found that the hard X-ray photon indices of both X-ray
binaries (XRBs) and AGNs are anti-correlated with the Eddington
ratios when the Eddington ratio is less than a critical value, while
they become positively correlated when the Eddington ratios higher
than the critical value (e.g., Wu and Gu 2008). This results provide
evidence for the accretion mode transition near the critical
Eddington ratio.

One of the uncertainties in {FR\, Is} is the respective contribution
of ADAFs and jets at various wavebands. The least controversial
nuclear emission are the radio and also optical emission, which is
believed to be dominated by the jet (e.g., Wu and Cao 2005; Chiaberge
\textit{et al} 2006). It is still an open question for the origin of the
core X-ray emission in {FR\,Is} (and also LLAGNs). One possibility
for the X-ray emission is dominated by the ADAF (e.g., Reynolds et
al. 1996b; Quataert \textit{et al} 1999; Merloni 2003). The other
possibility is come from the jet (e.g., Falcke \textit{et al} 2004; Markoff
\textit{et al} 2008), and recent observation on some individual extremely
low luminosity sources supported this possibility (e.g., Garcia et
al. 2005 for M\,31; Pellegrini \textit{et al} 2007 for NGC\,821, etc.). Then
an important question is \emph{systematically} in what kind of
condition the radiation from the jet or ADAF will be important in
X-ray band.

\section{Sample}
 The {FR\,I} sample used for the present investigation is selected
from Donato \textit{et al} (2004), which have the estimated black hole mass,
Bondi accretion rate, optical, radio, and X-ray nuclear emission
(Table 1). There are 9 {FR\,Is} in their sample which have compact
core X-ray emission
    and have been observed by $Chandra$. We excluded 3C\,270 since
    the optical emission may be obscured by its large intrinsic
    column density ($N_{H}\sim10^{22}\rm cm^{-2}$). Therefore, our final
    sample include 8 {FR\,Is}.

\section{Coupled accretion-jet model}
   The accretion flow is described by a hot, optically thin, geometrically thick advection-dominated accretion
   flow. Following the proposal of
Blandford and Begelman (1999), we parameterize the radial variation
of the accretion rate with the parameter $p_{\rm w}$ caused by the
possible wind (e.g., Stone \textit{et al} 1999), $\dot{M} =
\dot{M}_{\rm out}(R/R_{\rm out})^{p_{\rm w}}$, where $\dot{M}_{\rm
out}$ is the accretion rate at the outer boundary of the ADAF
$R_{\rm out}$. We calculate the global solution of the ADAF. The
viscosity parameter $\alpha$ and magnetic parameter $\beta$ (defined
as ratio of gas to total pressure in the accretion flow,
$\beta=P_{\rm g}/P_{\rm tot}$) are fixed to be $\alpha=0.3$ and
$\beta=0.9$. Another parameter is $\delta$, describing the fraction
of the turbulent dissipation which directly heats electrons.
Following Yuan \textit{et al} (2006), we use $\delta=0.3$ and $p_{\rm
w}=0.25$ in all our calculations. The radiative processes we
consider include synchrotron, bremsstrahlung and their
Comptonization. We set the outer boundary of the ADAF at the Bondi
radius $R_{B}=2GM_{\rm BH}/c_{s}^{2}$, where $c_{s}=\sqrt{\gamma
kT/\mu m_{p}}$ is the adiabatic sound speed of the gas at the Bondi
accretion radius, $T$ is the gas temperature at that radius,
$\mu=0.62$ is the mean atomic weight, $m_{p}$ is the proton mass,
and $\gamma=4/3$ is adiabatic index of the X-ray emitting gas. After
the ADAF structure is obtained, the spectrum of the flow can be
calculated (Yuan \textit{et al} 2003).

The jet model adopted in the present paper is based on the internal
shock scenario, widely used in interpreting gamma-ray burst (GRB)
afterglows (e.g., Piran 1999). A fraction of the material in the
accretion flow is assumed to be transferred into the vertical
direction to form a jet due to the possible standing shock near the
black hole (BH). From the shock jump conditions, we calculate the
properties of the post-shock flow, such as electron temperature
$T_{\rm e}$. The jet is assumed to have a conical geometry with
half-open angle $\phi$ and bulk Lorentz factor $\Gamma_{j}$ which
are independent of the distance from the BH. The internal shock in
the jet should occur as a result of the collision of shells with
different $\Gamma_{j}$, and these shocks accelerate a small fraction
of the electrons into a power-law energy distribution with index
$p$. We assume that the fraction of accelerated electrons in the
shock $\xi_{\rm e}=10\%$ and a typical value of $\phi=0.1$ (e.g.,
Laing \& Bridle 2002) in our calculations. Following the widely
adopted approach in the study of GRBs, the energy density of
accelerated electrons and amplified magnetic field are described by
two free parameters, $\epsilon_{\rm e}$ and $\epsilon_{\rm B}$.
Obviously, $\xi_{\rm e}$ and $\epsilon_{\rm e}$ are not independent.
The Lorentz factor of most {FR\,Is} are determined from the
observations and a typical value $\Gamma_{j}=2.3$ (corresponding to
$v/c=0.9$, e.g., Verdoes Kleijn \textit{et al} 2002; Laing \& Bridle 2002)
is adopted for those not having estimations, which will not affect
our main conclusion since that our results are not very sensitive to
the Lorentz factor.

\section{Results}
   We use the ADAF-jet model to fit the spectrum of {FR\,Is}.
 As we state in Sect.\ 1, the radio emission, and perhaps optical
as well, is from the jet. Based on the assumption to the jet model
described in Sect.\ 3, the contribution of the jet to the X-ray band
is well constrained once we require the jet model to fit the radio
and optical data. We then adjust the parameter of the ADAF and
combine it with the jet contribution to fit the X-ray spectrum.

\subsection{ADAF dominated the nuclear X-ray emission for higher Eddington ratio sources}
   The radio morphology and power of 3C\,346 would rank as either a
low-power {FR\,II} source or a high-power {FR\,I}. $Chandra$
observations have detected an unresolved core with $L_{X}(2-10\ \rm
keV)=1.9\times10^{43}\ \rm erg\ s^{-1}$ and a photon index of
$\Gamma=1.69\pm0.09$ (Donato \textit{et al} 2004). Figure 1 shows the
fitting result. The parameters of the jet are $\dot{m}_{\rm
jet}=3.5\times10^{-5}$, $\epsilon_{\rm e}=0.14$, $\epsilon_{\rm
B}=0.02$, and $p=2.4$. We find that the jet model can describe well
the radio and optical data. But the X-ray emission of the jet model
is several times lower than the $Chandra$ observations, and the
X-ray data can be well fitted by the ADAF. The required accretion
rate is $\dot{m}_{\rm out}=2.8\times10^{-2}$. The ratio of mass loss
rate in the jet to accretion rate of ADAF at 10 $R_{\rm S}$ is about
$\dot{m}_{\rm jet}/\dot{m}(10 R_{\rm S})=$0.9\%, where $R_{\rm S}$
is the Schwarzschild radius. This source is relatively luminous,
with $L_{\rm X}/L_{\rm Edd}=1.8\times10^{-4}$, and the X-ray
emission is dominated by the ADAF, not by the jet.
\begin{figure}[h]
\begin{minipage}[t]{0.45\linewidth}
\includegraphics[width=17pc]{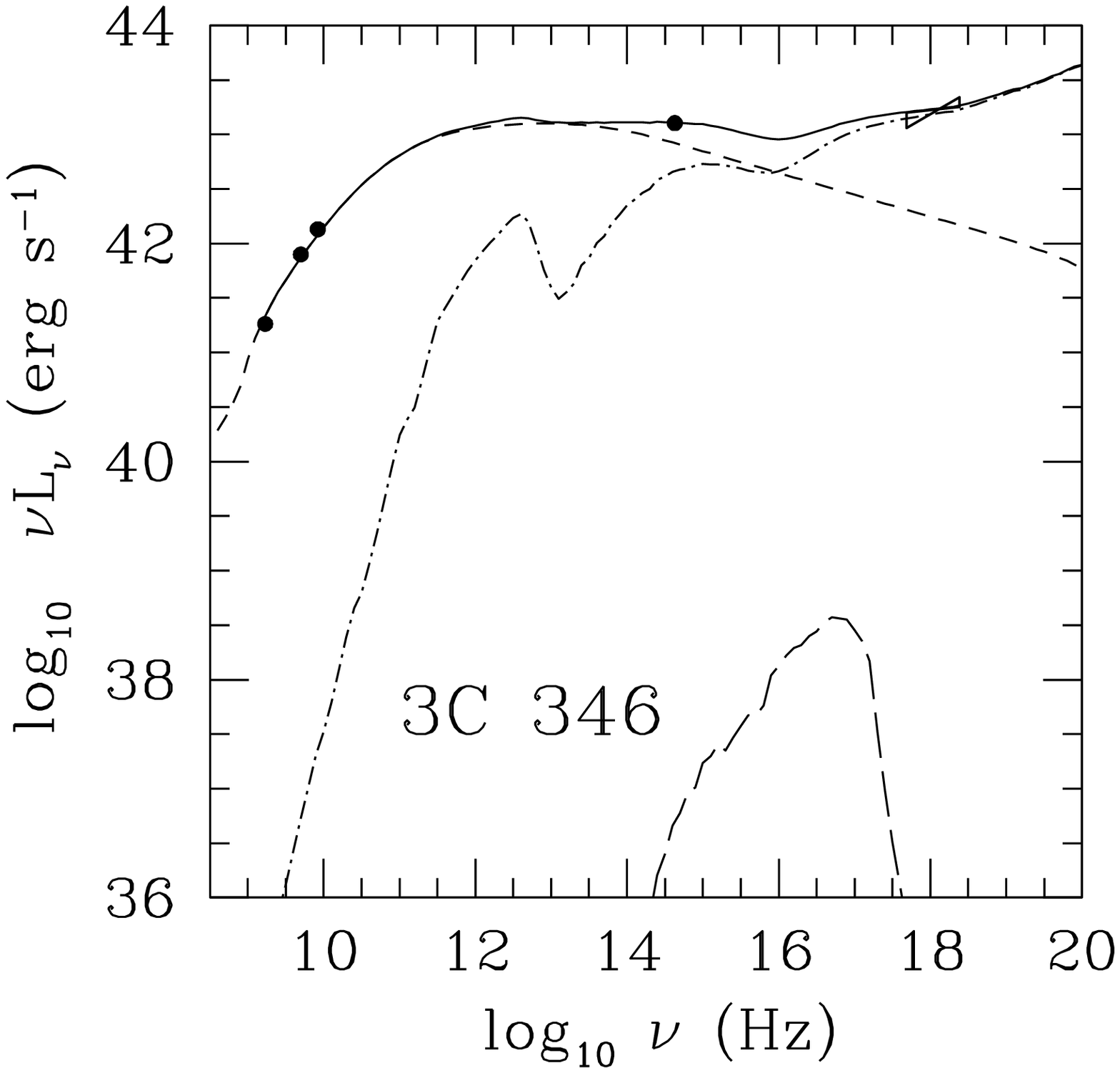}
\caption{\label{label}Spectral energy distribution of 3C\,346. The
dot-dashed, dashed, and the solid lines show the emissions from the
ADAF, jet, and their sum, respectively. The thin long-dashed line is
synchrotron-self-Compton spectrum of the jet.}
\end{minipage} \hspace{2pc}%
\begin{minipage}[t]{0.45\linewidth}
\includegraphics[width=17pc]{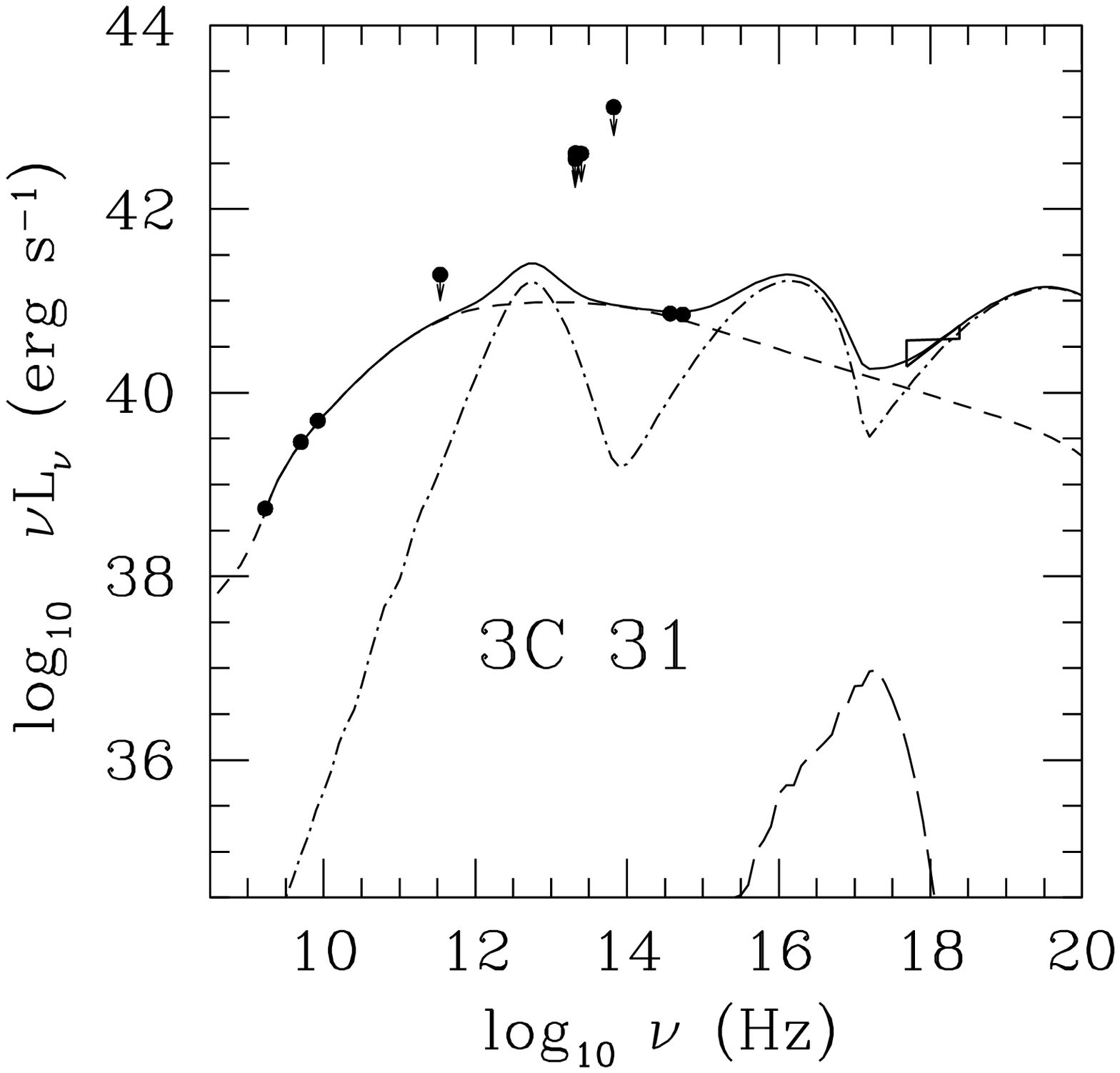}
\caption{\label{label}The same as Fig.\ 1, but for 3C\,31.}
\end{minipage}
\end{figure}

\subsection{ADAF+jet for moderate Eddington ratio sources}

 3C\,31 is a twin-jet {FR\,I} radio galaxy. The X-ray spectrum of the core is quite flat,
 with a photon index $\Gamma=1.48^{+0.28}_{-0.32}$ and $L_{X}(2-10\ \rm
keV)=4.7\times10^{40}\rm erg\ s^{-1}$ ($\sim4.4\times10^{-6}L_{\rm
Edd}$)(Evans \textit{et al} 2006).
   Figure 2 shows the fitting result of 3C\,31. The parameters of
the jet are $\dot{m}_{\rm jet}=2.7\times10^{-5}$, $\epsilon_{\rm
e}=0.2$, $\epsilon_{\rm B}=0.02$, and $p=2.5$. We find that the
radio, optical and even the soft X-rays nuclear emission (e.g., 
1\,keV) can be well-fitted by a pure jet model. However, the hard
X-rays cannot be fitted by a jet, but can be well-fitted by the ADAF
with an accretion rate of $\dot{m}_{\rm out}=3.7\times10^{-3}$. The
ratio, $\dot{m}_{\rm jet}/\dot{m}(10\ R_{\rm S})$, is about 9\%. We
can see that jet contribution is important at soft X-ray band, while
the ADAF contribution is important at the hard X-ray band.

This is also the case for other {FR\,Is} with moderate Eddington
ratios ($L_{X}/L_{\rm Edd}\sim several \times 10^{-6}$, e.g., 
3C\,317, B2\,0055+30, and possibly also 3C\,449), except B2\,0755+37, where
all the emission can be fitted by a pure jet model (see Wu \textit{et al}
2007 for more details).

\subsection{Jet dominated the nuclear X-ray emission for lower Eddington ratio sources}

The point-like nucleus is detected in 3C 66B by $Chandra$ yielding
$L_{X}(2-10\ \rm keV)=1.1\times10^{41}\rm ergs^{-1}$
($\sim1\times10^{-6}L_{\rm Edd}$) and a photon index of
$\Gamma=2.17^{+0.14}_{-0.15}$ (e.g., Donato \textit{et al} 2004). Figure 3
shows the fitting result of nucleus of 3C 66B. We find that all the
radio, sub-millimetre, optical and X-ray can be fitted by a pure jet
model very well (dashed line). The parameters of the jet are
$\dot{m}_{\rm jet}=1\times10^{-5}$, $\epsilon_{\rm e}=0.18$,
$\epsilon_{\rm B}=0.02$, and $p=2.35$. For illustration purposes, we
also show the X-ray emission using an ADAF model with $\dot{m}_{\rm
out}=2.6\times10^{-3}$ (dot-dashed line). The predicted spectrum by
an ADAF is too hard to be consistent with the observations (so the
accretion rate in the ADAF should be smaller than
$2.6\times10^{-3}$). The X-ray emission in this source should be
produced by the jet.

3C 272.1 is the lowest Eddington source ($\sim6.8\times10^{-8}L_{\rm
Edd}$), for which many waveband high resolution observations from
radio to X-ray have been carried out. Figure 4 shows the fitting
result of nucleus of 3C 272.1. The parameters of the jet are
$\dot{m}_{\rm jet}=4.9\times10^{-6}$, $\epsilon_{\rm e}=0.28$,
$\epsilon_{\rm B}=0.005$, and $p=2.5$. We can see that the radio,
sub-millimetre, optical, and especially, the X-ray emission can be
fitted by the jet model very well. On the other hand, the predicted
spectrum by an ADAF (with $\dot{m}_{\rm out}=1.9\times10^{-3}$) is
too hard to be consistent with the observations. So in this source,
the X-ray emission is also dominated by the jet.

\begin{figure}[h]
\begin{minipage}[t]{0.45\linewidth}
\includegraphics[width=17pc]{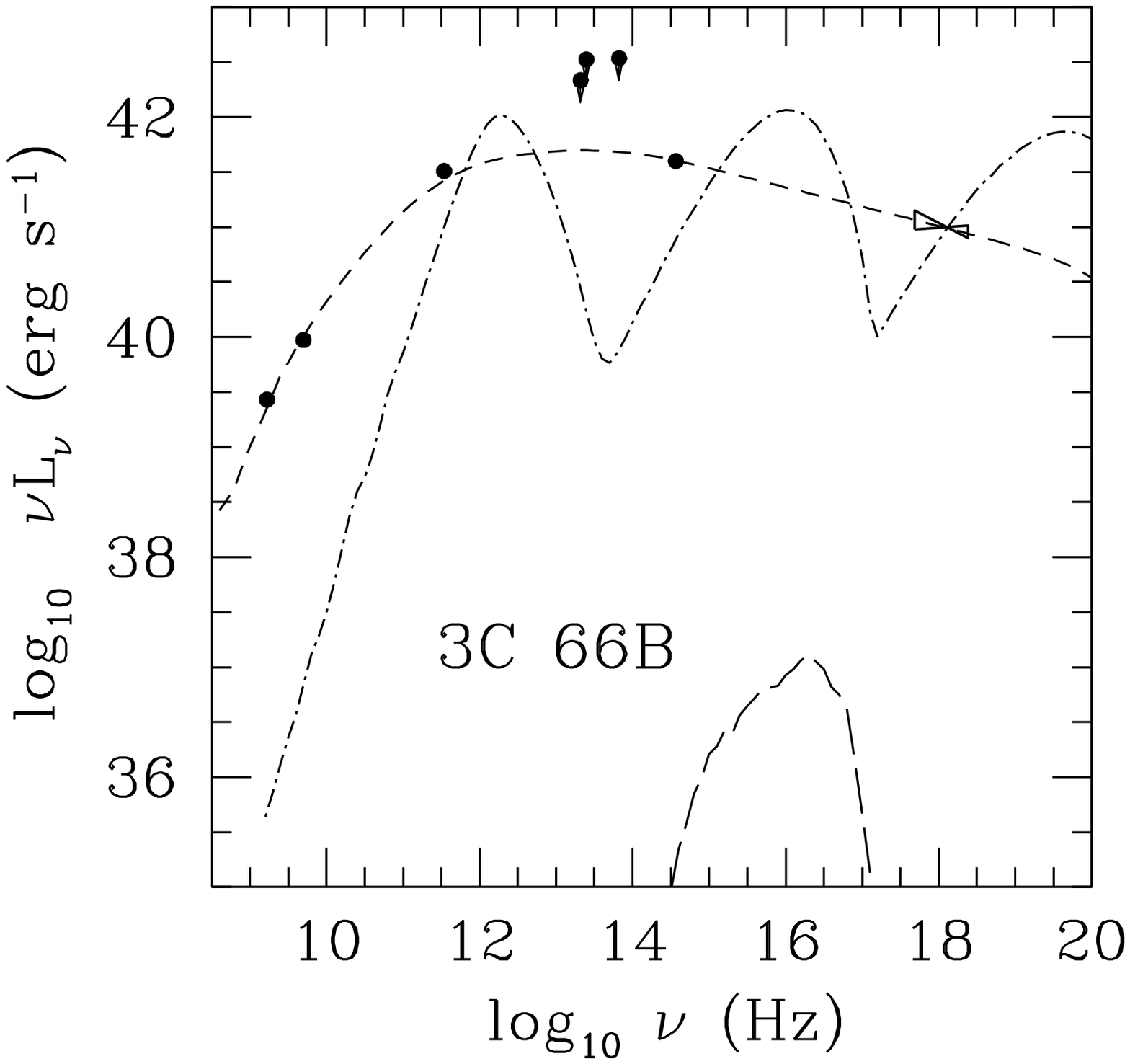}
\caption{\label{label}The same as Fig.\ 1, but for 3C 66B.}
\end{minipage} \hspace{2pc}%
\begin{minipage}[t]{0.45\linewidth}
\includegraphics[width=17pc]{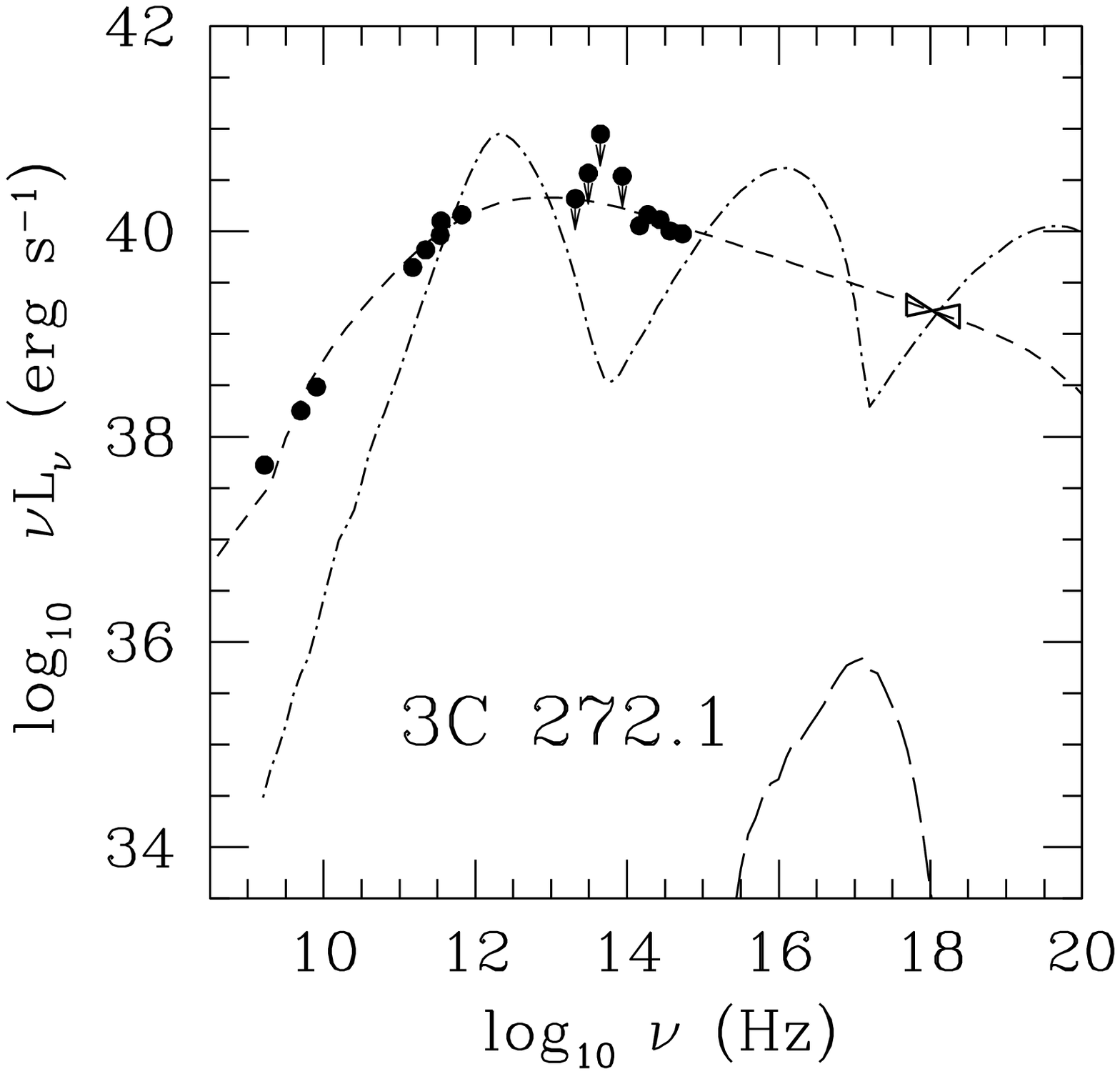}
\caption{\label{label}The same as Fig.\ 1, but for 3C 272.1.}
\end{minipage}
\end{figure}

\subsection{Fuel supply in these {FR\,Is}}

Typically, the Bondi accretion rate is a good estimation to the mass
accretion rate. However, Pellegrini (2005) show that there is no
relation between the nuclear X-ray luminosity and the Bondi
accretion rate in LLAGNs, and the X-ray emission of some sources is
higher than the values predicted by ADAFs with the Bondi accretion
rate. The Bondi accretion rate in our sample has been estimated
(see, Donato \textit{et al} 2004). We find that the accretion rates
$\dot{m}_{\rm out}$ required in our model of four {FR\,Is} (3C 346,
3C 31, 3C 449, and 3C 317), among the five in which we can have good
constraints to their accretion rates, are higher than their Bondi
rates by factors of 9, 18, 112, and 1.05, respectively (see also
table 1). Given that the radial velocity of the accretion flow is
$\alpha c_s$, a more accurate estimation of the accretion rate is
$\dot{m}_{\rm out}\sim\alpha\dot{m}_{\rm Bondi}$ where $\alpha$ is
the viscous parameter (Narayan 2002). Therefore, the Bondi accretion
rate is only a lower limit of the real rate and other fuel supply
must be important, such as the gas released by the stellar
population inside the Bondi radius (Soria \textit{et al} 2006; Pellegrini
2007).

\begin{table}[h]
\caption{\label{ex}Accretion and jet properties.}
\[
\centering \resizebox{\textwidth}{!}{%
\begin{tabular}{lcccccccc}
\br Source & $\rm log10 (\frac{M_{\rm BH}}{\rm M_{\odot}})$ &
$\frac{L_{\rm X}(2-10\rm keV)}{L_{\rm Edd}}$ & $\frac{\dot{m}_{\rm
jet}}{\dot{m}(10R_{\rm S})}$ & $\dot{m}(R_{\rm B})^{a}$ &
$\dot{m}_{B}^{b}$ & $\frac{L_{\rm Kin}}{\dot{M}(10R_{\rm S})c^{2}}$\\
\mr
3C 346       & 8.89   &  $1.8\times10^{-4}$        & 0.91\%       & $2.8\times10^{-2}$   & $3.1\times10^{-3}$  &   0.03 \\
B2 0755+37   & 8.93   &  $5.2\times10^{-6}$        & $>3.9\%$     & $<8.9\times10^{-3}$  & $3.2\times10^{-2}$  &   $>0.10$\\
3C 31        & 7.89   &  $4.4\times10^{-6}$        & 9.0\%        & $3.7\times10^{-3}$   & $2.1\times10^{-4}$  &   0.19\\
3C 317       & 8.80   &  $3.4\times10^{-6}$        & 8.9\%        & $4.7\times10^{-3}$   & $4.5\times10^{-3}$  &   0.28\\
B2 0055+30   & 9.18   &  $2.4\times10^{-6}$        & 3.5\%        & $2.7\times10^{-3}$   & $1.4\times10^{-2}$  &   0.08\\
3C 66B       & 8.84   &  $1.0\times10^{-6}$        & $>5.9\%$     & $<2.6\times10^{-3}$  & $2.5\times10^{-2}$  &   $>0.18$\\
3C 449       & 8.42   &  $8.0\times10^{-7}$        & 14.3\%       & $1.9\times10^{-3}$   & $1.7\times10^{-5}$  &   0.44\\
3C 272.1     & 8.35   &  $8.3\times10^{-8}$        & $>7.3\%$     & $<1.9\times10^{-3}$  & $6.0\times10^{-2}$  &   $>0.22$\\
\br
\end{tabular}
}
\]
 (a) $\dot{m}(R_{\rm B})$ is the
dimensionless accretion rate at the Bondi radius through our spectra
fitting; (b) $\dot{m}_{B}$ is the dimensionless Bondi accretion rate
estimated from the X-ray observation.
\end{table}

\section{Discussion and Conclusion}
  Two possibilities exist for the X-ray origin in {FR\,Is}: ADAF or jet.
We use a coupled ADAF-jet model to fit the multi-waveband spectrum to
try to investigate this problem. We find that the jet can well
describe the radio and optical spectra for most {FR\,Is} in our
sample. The soft X-ray flux at $\sim$1\,keV of all {FR\,Is} is
roughly consistent with the predictions of the jet. This result
indicates why a tight correlation is found among radio, optical and
soft X-ray (Evans \textit{et al} 2006; Balmaverde \textit{et al} 2006). For 3C 346,
the highest X-ray luminosity source in our sample which has $L_{\rm
X}=1.8 \times 10^{-4} L_{\rm Edd}$, its X-ray spectrum is dominated
by the ADAF and the jet contribution is negligible. However, for the
four sources with ``intermediate X-ray luminosities'' ({B2\,
0755+37}, 3C 31, 3C 317, and B2\,0055+30; $L_{\rm X}=(2.4-5.2) \times
10^{-6} L_{\rm Edd}$), their X-ray origin is complicated. The X-ray
spectrum of {B2\,0755+37 is dominated by the jet, while for 3C 31 and
B2\,0055+30} spectra are dominated by the ADAF. For the other one (3C
317), the contributions of the ADAF and jet are comparable. For the
three least luminous sources (3C 66B, 3C 449, and 3C 272.1) which
have $L_{\rm X}=(6.8\times 10^{-8}-1\times 10^{-6}) L_{\rm Edd}$,
their X-ray spectra are dominated by the jet. The X-ray emission of
3C 449 is also interpreted by the sum of a jet and an ADAF, which
requires higher quality data to further constrain it. Our results
are roughly consistent with the predictions of Yuan and Cui (2005).
The ``intermediate luminosity'' here in our sample corresponds to
the critical luminosity in Yuan and Cui (2005). However, the former
is about 10 times higher than the latter. The value of the critical
luminosity depends on the ratio of the mass loss rate in the jet to
the mass accretion rate in the ADAF. This ratio is adopted in Yuan
and Cui (2005) from fitting the data of a black hole X-ray
binary---XTE\,J1118+480. The current result indicates that the ratio
in FR Is is about 10 times higher than in XTE\,J1118+480. One
possible reason could be that systematically the black holes in FR
Is are spinning more rapidly. We note that the critical luminosity
(or Eddington ratio) in this work is for the transition of
\emph{X-ray emission} dominated by the ADAF or by the jet, which is
different from the critical value for the transition of the
\emph{jet power dominated systems} and a\emph{ccretion power
dominated systems} (Wu and Cao 2008 for more details and references
therein).

The kinetic luminosity, $L_{\rm kin}=\Gamma_{j}(\Gamma_{j}-1)
\dot{M}_{\rm jet}c^{2}$, can be derived from our modeling results.
We use $\eta_{\rm jet}=L_{\rm kin}/\dot{M}(10R_{S})c^{2}$ to
describe the efficiency of the jet power converted from the
accretion power, where $\dot{M}(10R_{S})$ is mass accretion rate at
$10R_{S}$. We find that $\eta_{\rm jet}=0.03-0.44$ for the sources
in this sample (see Table 1), and most of them (six of eight) have
$\eta_{\rm jet}$ significantly higher than 0.057, which is the
largest available accretion energy at the innermost stable circular
orbit (ISCO) for a non-rotating black hole. It implies that the
black holes in these sources are spinning rapidly (so ISCO is
smaller thus more accretion energy is available).

 \ack Wu Q W thanks the local organizing
committee of the conference for partial financial support. This work
is supported by the One-Hundred-Talent Program of China, the
National Science Fund for Distinguished Young Scholars (grant
10325314), the NSFC (grants number 10773024, 10773020, 10703009 and
10633010), and a postdoctoral fellowship of the KASI.

\section*{References}
\medskip
\begin{thereferences}
\item Balmaverde B, Capetti A and Grandi P 2006 {\it A\&A} {\bf 451} 35
\item Bicknell G V 1995 {\it ApJS} {\bf 101} 29
\item Blandford R D and Begelman M C 1999 {\it MNRAS} {\bf 301} L1
\item Chiaberge M, Gilli R, Macchetto F D and Sparks W B 2006 {\it ApJ} {\bf 651} 728
 \item Donato D, Sambruna R M and  Gliozzi M 2004 {\it ApJ} {\bf 617} 915
 \item Evans D A, Worrall D M, Hardcastle M J, Kraft R P and Birkinshaw M 2006 {\it ApJ} {\bf 642} 96
\item Falcke H, K$\ddot{\rm o}$rding E and Markoff S. 2004 {\it A\&A}
        {\bf 414} 895
\item Fanaroff B L and Riley J. M. 1974 {\it MNRAS} {\bf 167} 31
\item Garcia M R \textit{et al} 2005 {\it ApJ} {\bf 632} 1042
\item Hardcastle M J, Evans D A and Croston J H 2007 {\it MNRAS}
{\bf 376} 1849
\item Ho L C 2008 {\it ARA\&A} ({\it Preprint:} astro-ph/08032268)
\item Laing R A and Bridle A H 2002 {\it MNRAS} {\bf 336} 328
\item Markoff S. \textit{et al} 2008 {\it ApJ} {\bf 681} 905
\item Merloni A, Heinz S and di Matteo T 2003 {\it MNRAS} {\bf 345} 1057
\item Narayan R and McClintock J E 2008 {\it New Astronomy Reviews}
      {\bf 51} 733
\item Narayan R 2002 \textit{Lighthouses of the Universe} ed M Gilfanov \textit{et al} (Berlin: Springer) 405
\item Pellegrini S \textit{et al} 2007 {\it ApJ} {\bf 667} 749
\item Piran T 1999 {\it PhR} {\bf 314} 575
\item Quataert E, di Matteo T, Narayan R and Ho L C 1999 {\it ApJ} {\bf 525} L89
\item Reynolds C S, di Matteo T, Fabian A C, Hwang U and Canizares C R 1996a {\it MNRAS} {\bf 283} 111
\item Reynolds C S, Fabian A C, Celotti A and Rees M J 1996b {\it MNRAS} {\bf 283} 873
\item Soria R \textit{et al} 2006 {\it ApJ} {\bf 640} 143
\item Stone J M, Pringle J E and Begelman M
       C 1999 {\it MNRAS} {\bf 310} 1002
\item Verdoes Kleijn G A, Baum S A, de Zeeuw P T and O'Dea C P 2002 {\it AJ} {\bf 123} 1334
\item Wu Q W and Cao X W 2008 {\it ApJ} ({\it Preprint:} astro-ph/08072288)
\item Wu Q W and Gu M F 2008 {\it ApJ} {\bf 682} 212
\item Wu Q W, Yuan F and Cao X W 2007 {\it ApJ} {\bf 669} 96
\item Wu Q W and Cao X W 2005 {\it ApJ} {\bf 621} 130
\item Yuan F 2007, \textit{ASP Conf. Ser.} {\bf 373} 95
\item Yuan F, Shen Z Q and Huang L 2006 {\it ApJ} {\bf 642} 45
\item Yuan F, and Cui W 2005 {\it ApJ} {\bf 629} 408
\item Yuan F, Quataert E and Narayan R  2003 {\it ApJ} {\bf 598} 301

\end{thereferences}
\end{document}